\documentclass[aps,pre,twocolumn,groupedaddress,showpacs]{revtex4}
\usepackage{graphicx}
\usepackage{amssymb}

\begin{document}

\title{Avalanche statistics and time-resolved grain dynamics for a driven heap}
%Instantaneous grain dynamics in continuous and avalanche flows
%Instantaneous dynamics and statistics of granular flows

\author{A.R. Abate, H. Katsuragi,\footnote{Permanent address: Department of
Applied Science for Electronics and Materials, Kyushu
University, Fukuoka 816-8580, Japan} and D.J. Durian} \affiliation{Department
of Physics and Astronomy, University of Pennsylvania,
Philadelphia, PA 19104-6396, USA}

\date{\today}

\begin{abstract}
We probe the dynamics of intermittent avalanches caused by steady addition of grains to a quasi-two dimensional heap. To characterize the time-dependent average avalanche flow speed $v(t)$, we image the top free surface. To characterize the grain fluctuation speed $\delta v(t)$, we use Speckle-Visibility Spectroscopy. During an avalanche, we find that the fluctuation speed is approximately one-tenth the average flow speed, $\delta v \approx 0.1 v$, and that these speeds are largest near the beginning of an event. We also find that the distribution of event durations is peaked, and that event sizes are correlated with the time interval since the end of the previous event.  At high rates of grain addition, where successive avalanches merge into smooth continuous flow, the relationship between average and fluctuation speeds changes to $\delta v \sim v^{1/2}$. \end{abstract}


\maketitle

%=========================================================================================
\section{Introduction}

The behavior of granular media continues to pose significant challenges to both theory and experiment \cite{Jaeger1996, Duran, LiuNagelBook, GDRMiDi04, Aranson2006}.  One source of the richness in this field is that flows occur only after the forcing exceeds a threshold.  The flow response is thus highly nonlinear, and cannot be understood as a small excitation above the static state.  For example, grains in a hopper remain at rest unless the outlet underneath is greater than a few grain diameters.  Grains on a vibrated plate remain at rest unless the peak acceleration is greater than $g$.  And grains on the surface of a pile remain at rest unless the slope is greater than an angle of maximum stability.  In all cases, the resulting flows can be smooth and hydrodynamic-like at very high forcing, or temporally intermittent at low forcing just beyond threshold.

In the case of surface flows, granular avalanches have been observed in a variety of geometries including a rotating drum \cite{Jaeger1989, Rajchenbach1990, Evesque91, Rajchenbach2002, duPont2005}, an inclined plane \cite{Daerr99, Borzsonyi2005}, and a heap to which grains are added \cite{Jaeger1989, Frette96, Makse97, PierrePRL00, PierreAO01}.  Much effort concerns the size of the avalanches and the shape of the coarse-grained (a.k.a. hydrodynamic) velocity profiles.  But what is the nature of the grain-scale dynamics? Fluctuations of individual grains away from the average velocity lead to inelastic collisions that dissipate energy and that ultimately control the nature and rate of the flows.  While the microscopic grain dynamics is hence crucial to a fundamental understanding, it is particularly difficult to measure in slow dense flows because the mean-free path and mean-free time for grain-grain collisions can be too short for high-speed video, which only captures surface behavior anyway.  While diffusing-wave spectroscopy (DWS) is capable of measuring grain-scale fluctuations within the bulk, even for collision rates as high as $10^5$ per second and for mean free paths as short as $10^{-4}$ times the grain size \cite{Menon1997, Menon1997PRL}, it requires that the dynamics be time-independent \cite{Weitz1993, DWSrev97, PierrePRE98}.  For time-dependent avalanching flows, DWS must be extended by consideration of higher-order intensity correlations because the electric field statistics become non-Gaussian \cite{PierrePRL00, PierreAO01}.  However, interpretation of such data relies on the assumption that flows start and stop instantaneously, such that the average and fluctuation speeds remain constant during the course of an event.

In this paper we apply a time-resolved multi-speckle dynamic light scattering technique known as speckle-visibility spectroscopy (SVS) \cite{Dixon2003, Bandyopadhyay2005} to avalanches on a granular heap confined between two parallel plates.  This method is applicable to slow dense flows, just like DWS and the higher-order extensions, but is capable of resolving the evolution of dynamics throughout the course of an avalanche event.  Thus we measure in detail not just the statistics of avalanche sizes, but also their dynamics.  And by comparison of the resulting fluctuation speeds with coarse-grained average speeds found by high-speed video, we find that the nature of microscopic dynamics is different during an avalanche than during continuous flow.

%=========================================================================================
\section{Experimental methods}

\subsection{Granular system}

A quasi-two-dimensional heap is created by steady addition of grains between two parallel vertical walls, made of static-dissipating Lucite plates closed at bottom and along one vertical edge, as in Refs.~\cite{PierrePRL00, PierreAO01}.  The granular medium is dry, cohesionless glass beads with diameter range $0.25-0.35$~mm, repose angle $\theta_r=25^o$, and density $\rho=1.5~{\rm g/cm}^3$.  The inner wall dimensions are $28.5\times 28.5$~cm$^2$, so that the length along the slope of the heap is $28.5~{\rm cm}/\cos\theta_r=31~{\rm cm}$.  The channel width is $w=9.5$~mm, equal to approximately thirty grains across.  The flux $Q$ of grains onto the heap may be varied widely in increments of 0.005~g/s, and be held constant for extended periods, as follows.  A large funnel is positioned above the back of the channel, filled with grains, and allowed to drain under gravity.  The output is connected to a valve, which uses a moveable knife-edge to divert the desired flux onto the heap and to expunge the rest into a storage bucket for manual recycling.  The free-fall of grains onto the top of the heap is broken by an Aluminum bar, sandwiched between the plates to form a funnel with 1.5~cm outlet 3~cm above the top of the heap.

Several distinct flow regimes are observed depending on the flux $Q$ of grains. Above a critical flow rate of $Q_c=0.36$~g/s, the same value as in Ref.~\cite{PierrePRL00}, the flow along the surface of the heap is smooth and continuous \cite{Khakhar2001, Komatsu2001, Jop2005}.  Below $Q_c$, the flow is intermittent, accomplished by a series of discrete avalanches that start at the top of the heap.  About every third or fourth avalanche is a large, system-wide event in which the entire surface flows and expunges sand out the bottom of the channel; after such an event, the surface of the heap is smooth with a constant slope along its entire length.  Other avalanches are smaller and stop before reaching the bottom; after these events, the surface of the heap is uneven, with discrete steps between long regions of constant slope.  Very far below $Q_c$, the behavior is quasi-static in that successive avalanches are independent of one another and also in that variation of $Q$ affects the time between events but not the dynamics during flow.  As $Q_c$ is approached from below, successive avalanches merge to an extent set by the value $Q_c-Q$.  For the avalanche experiment reported below, the flux of grains is held fixed at $Q=0.07$~g/s.  This lies in the quasi-static regime, but only slightly below the point at which successive events start to merge in order to maximize the number of events observed per unit runtime.  At the very end of the paper, grain dynamics during avalanche are compared with those under continuous flow at grain fluxes across the range $Q_c=0.36~{\rm g/s}<Q<3.3~{\rm g/s}$.

Before describing the diagnostic tools and their application to behavior near the top free surface, we first remark upon the three-dimensional character of the velocity profiles in the continuous flow regime.  Using particle imaging velocimetry, as described below, we find that the flow speed at the side walls decreases nearly exponentially with depth.  The decay length is comparable to the channel width.  We also find that the flow profile is approximately parabolic across the top free surface, but with wall slip such that the speed at the center is 1.4 times faster than the speed at the wall.  For a known flux of 2.5~g/s, multiplying these forms and integrating over the cross section of the heap gives a flux estimate of 2.2~g/s.  The good agreement suggests that flow along the top and side surfaces is representative of flow within the bulk.

\subsection{Diagnostics}

Next we describe technical details for two methods for characterizing time-resolved avalanche dynamics.  Both are applied to the top surface of the granular heap, over the region between 2 and 4~cm from the outlet as measured along the slope.  The only avalanches that pass through this field of view are large ones, which span the entire surface of the heap.  The first diagnostic is a variant of Particle-Image Velocimetry (PIV) \cite{Adrian91, AdrianEF05}, which gives $v(t)$ the time-dependent coarse-grained hydrodynamic surface flow speed.  While originally developed for fluids, this method is well suited for granular media, as in eg. Refs.~\cite{Medina98, Lueptow2000, Tischer2001, Pudasaini2005}.  The second diagnostic is Speckle-Visibility Spectroscopy (SVS)~\cite{Dixon2003, Bandyopadhyay2005}, which gives $\delta v(t)$ the time-dependent speed of microscopic grain-scale fluctuations around the hydrodynamic flow.  These methods are implemented one at a time using an 8~bit Basler digital linescan camera, with $2\times1024$ pixels and 58~kHz maximum frame rate, controlled using LabVIEW v7.1 and the National Instruments Vision Toolkit.  Images are continuously captured and processed, so that raw video data need not be saved and so that extremely long duration runs are possible.

\subsubsection{Particle imaging}

\begin{figure*}
\includegraphics[width=6.5in]{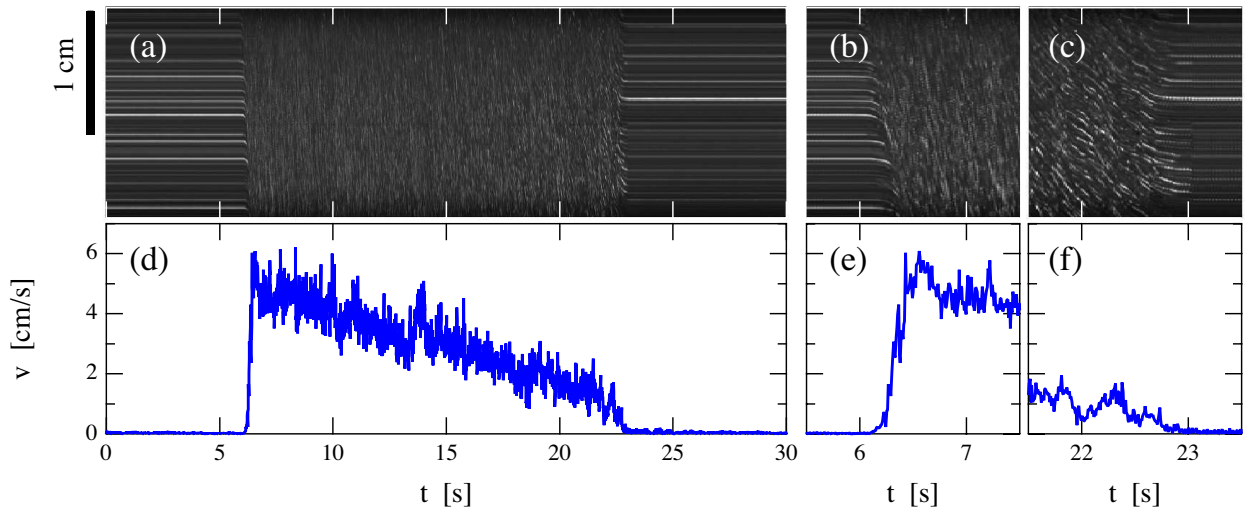}
\caption{(Color online) Time-stacked linescan images of the surface of the heap for an avalanche (a) and for zoom-ups of it turning on (b) and turning off (c). Sand grains appear as bright lines with slope proportional to their speed down the channel. The slope is measured by computing the maximum of the cross-correlation function for consecutive time snapshots. The bottom row shows the flow speeds extracted from the corresponding images above. }\label{vraw}
\end{figure*}

The time-dependent average speed, $v(t)$, of grains at the top free surface may be deduced from the spatial cross-correlation of successive images, as follows. Here, a bright halogen lamp is placed about 1~m away from the heap, shining down between the Lucite plates.  The linescan camera is placed about 20~cm away, with optical axis normal to the heap and and with the two rows of pixels oriented parallel to the flow direction.  The camera is fitted with a lens, such that the field of view is a 2~cm long strip, $39~\mu$m wide, located half way between the side walls.  Under these conditions, surface beads reflect light back to the camera and appear as a dark circles with central bright spots with object size of about 1/4 bead diameter and image size of size 4 pixels.  The frame rate of the camera is set to 1~kHz, such that for a typical flow speed of 4~cm/s the beads move about 2.5 pixels per frame.

While beads are thus imaged, it is not necessary to identify and track their positions individually.  Rather, the ensemble average speed of all imaged beads is found by the average displacement between successive frames.  This displacement is computed as the peak position of the spatial cross-correlation of successive images.  Here, correlations are found by Fourier methods, and the peak positions are identified by parabola fit to the cross-correlation function.

Example image data and flow speeds for a typical avalanche are shown in Fig.~\ref{vraw}, including a blow-up of the beginning and end of the event.  The top row shows space-time plots of raw grayscale images.  Before and after the event, when grain are at rest, the individual bright spots all remain at the same location and hence cause horizontal streaks in the plot.  During the event, the bright spots move and hence cause streaks with slope set by individual grain speeds.  The average streak slope, computed as per above, gives the ensemble-average grain speed shown in the bottom row.  Since grains diffuse laterally during flow, individual streaks last for only a finite time duration.

Several points are to be noted in Fig.~\ref{vraw}. First, the grain speeds are not constant during the avalanche.  Rather, $v(t)$ increases to a maximum at the beginning of the event and then gradually decreases back to zero. A front of flowing sand is observed to sweep through the field of view in part (b); this sets the rise time of $v(t)$ rather than the acceleration of grains from rest.  By contrast, grains in the field of view in part (c) all appear to come to rest at the same time.  The final approach to rest is very abrupt, nearly but not quite discontinuous. The scale of noise in $v(t)$ data is consistent with the estimate $\Delta v = f/(M\sqrt{N}) = 0.5~{\rm cm/s}$, where $f$ is the frame rate, $M$ is the magnification in pixels per cm, and $N$ is the number of beads in the field of view.

\subsubsection{Speckle visibility}

\begin{figure*}
\includegraphics[width=6.5in]{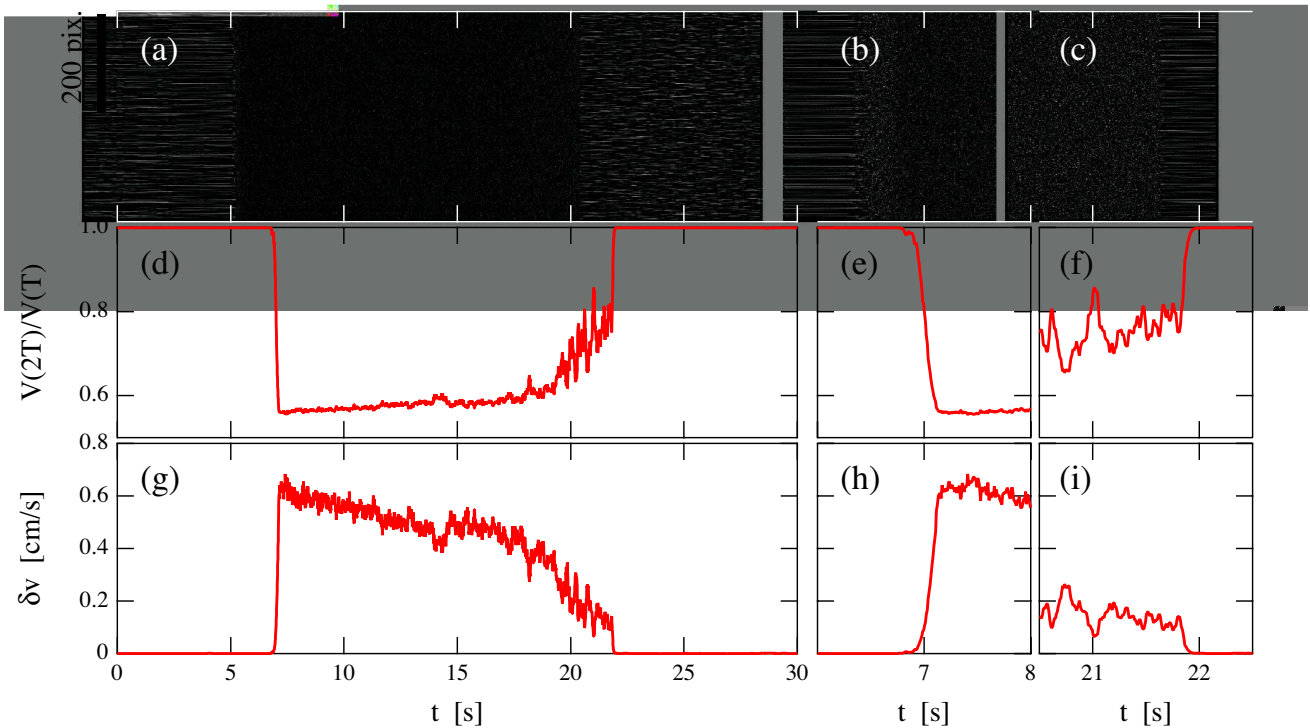}
\caption{(Color online) Time-stacked line scan images of the speckle pattern for an avalanche (a) and zoom-ups of it turning on (b) and turning off (c).  The middle row shows the corresponding variance ratios, which are transformed into the fluctuation speeds in the bottom row.}\label{dvraw}
\end{figure*}

The time-dependent fluctuation speed, $\delta v(t)$, of grains at the top free surface may be deduced from the the visibility of laser speckle formed by backscattered light.  This method has been dubbed Speckle-Visibility Spectroscopy (SVS)~\cite{Bandyopadhyay2005}, and has been applied to grains subject to periodic vibration~\cite{Dixon2003}, to colloids after cessation of shear~\cite{IanniPRE2006}, and to foams subject to coarsening~\cite{Bandyopadhyay2005}.  Closely related methods are laser-speckle photography \cite{BriersOC81} and Time-Resolved Correlation (TRC) \cite{wongRSI, pineRSI02, lucaTRC}.  Here, coherent light from a Nd:YAG laser, wavelength $\lambda=532$~nm, power 4~W, is expanded to a Gaussian diameter of 1.3~cm and is directed normally onto the heap at the same location that average speeds were observed.  An aperture blocks the beam near the channel walls, so that the illumination spot is roughly $0.95\times 1.3~{\rm cm}^2$.  Incident photons diffuse in the medium with a transport mean free path of a few grains. Therefore the typical photon emerges from the sample after a couple scattering events from grains close to the surface~\cite{Cox01}.  The linescan camera is placed about 15~cm away, with optical axis normal to the heap.  Now, however, the lens is removed and is replaced by a laser line filter to eliminate room light.  Under these non-imaging conditions, the backscattered produces a speckle pattern in the plane of the camera such that the speckle size is about 0.7~pixels.  As grains move relative to one another, the speckle pattern fluctuates and hence appears visible only if the camera frame rate is fast compared to the speckle lifetime, which is set roughly by the time $\lambda/\delta v$ for adjacent scattering sites to move one wavelength apart.  Note that for uniform translation of the sample, with no motion of grains relative to one another, the speckle pattern changes much more slowly over the time $L/v$ required for a new ensemble of grains to come into the field of view $L$.  Hence the idea of SVS is to deduce the grain fluctuation speed $\delta v$ from the visibility of the speckle for a given exposure duration $T$.  Here the camera is operated at maximum frame rate, 58~kHz, giving $T=17.24~\mu{\rm s}$.  For convenience, only the central $2\times430$ pixels are processed.  The laser intensity is adjusted so that the average grayscale level is 50.

The visibility of the speckle may be quantified by the variance of intensity levels, $V(T)\propto \langle I^2\rangle_T - \langle I\rangle^2$, where $\langle\cdots\rangle_T$ denotes the average over pixels exposed for duration $T$.  Note that the average intensity is independent of $T$, so no subscript is placed on $\langle I\rangle$.  The proportionality constant of $V(T)$ is set by the laser intensity and the ratio of speckle to pixel size.  It may be neatly canceled by considering the variance ratio $V(2T)/V(T)$, where the numerator is found from a ``synthetic exposure'' equal to the sum of successive images.  For diffusely backscattered light from particles moving with random ballistic motion, the theory of SVS \cite{Dixon2003,Bandyopadhyay2005} gives the variance ratio as
\begin{equation}
{V(2T)\over V(T)} = {e^{-4x}-1+4x \over 4\left(e^{-2x}-1+2x\right)}
\label{svsratio}
\end{equation}
where $x=(4\pi \delta v/\lambda)T$.  This equation may be numerically solved for $\delta v$ vs time in terms of data for variance vs time.  A rational approximation that has the correct small and large $x$ limits of $1 - 2x/3 + O(x^2)$ and $(4+1/x)/8 + O(1/x^2)$, respectively, and that may be inverted analytically for a good initial guess, is given by $V(2T) / V(T) \approx (1+7x/6+x^2) / (1+11x/6+2x^2)$.

Example speckle images and SVS analysis for a typical avalanche are shown in Fig.~\ref{dvraw}, including a blow-up of the beginning and end of the event.  The top row shows space-time plots of raw grayscale images of the speckle pattern.  Before and after the event, when grain are at rest, the speckles are nearly static and hence appear as horizontal streaks.  The variance of grayscale levels is nearly maximum, both as measured over exposure times $T$ and $2T$.  Hence the variance ratio is nearly one and the random grains speeds are nearly zero, as shown in the second and third rows.  However the speckle streaks do not extend indefinitely in the top row of Fig.~\ref{dvraw}, with speckles lasting on the order of a few seconds.  Over this time the grains experience wavelength-scale motion, due perhaps to ambient vibration, to the addition of grains at the top of the heap, or even to thermal expansion and contraction \cite{SchifferNature06}.  The motion during the avalanche is much faster, so that the random bright-dark pattern of speckles becomes a washed-out blur readily visible only near the beginning and end of the event.  Therefore the variance of grayscale levels decreases, more so for increasing exposure duration, as quantified by the variance ratio $V(2T)/V(T)$ displayed in the middle row of Fig.~\ref{dvraw}.  Using Eq.~(\ref{svsratio}), this data gives the random fluctuation speeds $\delta v$ displayed in the bottom row.  Note that, like the average speed $v$, the fluctuation speed is maximum at the beginning of the event, then it decreases gradually and finally halts abruptly.  Also note that the two speed scales differ by a factor of roughly ten, $\delta v=O(0.1)v$.

%=========================================================================================

\section{Results}

\begin{figure}
\includegraphics[width=3.0in]{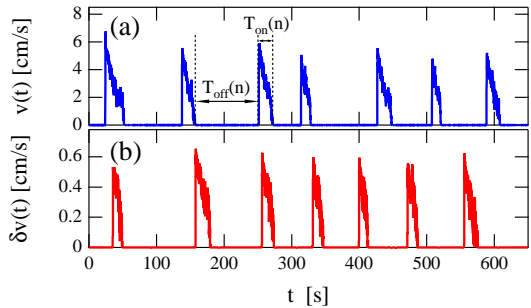}
\caption{(Color online) (a) Average and (b) fluctuation speed timetraces for independent observations with the same sand flux $Q=0.07$ g/s.  The on and off times, $T_{\rm on}$ and $T_{\rm off}$, for the $n^{\rm th}$ avalanche event are defined as shown.} \label{vtdvt}
\end{figure}

Example ten-minute time-traces of the average flow speed $v(t)$, from imaging, and of the fluctuation speed $\delta v(t)$, from SVS, are displayed in Fig.~\ref{vtdvt}.  Each spike represents one avalanche, in which both speeds suddenly rise from zero, persist briefly, and then vanish.  Note that successive avalanches are roughly similar in size, duration, and spacing. In this section we analyze two 37~hour runs, spanning 1436 events in $v(t)$ and 1472 events in $\delta v(t)$.  We find that the root-mean-squared speeds during flow are $\sqrt{\langle v^2\rangle} = 3.0$~cm/s and $\sqrt{\langle \delta v^2\rangle} = 0.37$~cm/s, and that the fraction of time spent in flow and at rest are $f_1=0.2$ and $f_0=0.8$, respectively.  First we consider the times $T_{\rm on}$ and $T_{\rm off}$, defined in Fig.~\ref{vtdvt}, for the duration of events and for the intervening quiescent periods.  Then we consider switching functions that describe the probability for the heap to change between flow and rest states after a given time delay.  Lastly we consider the dynamics of an average event and the relation between the instantaneous average and fluctuation speeds.

\subsection{Event durations}

\begin{figure*}
\includegraphics[width=6.5in]{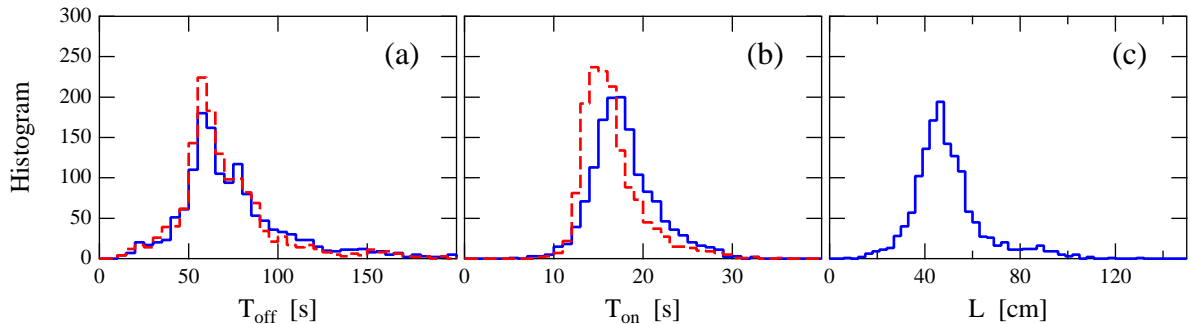}
\caption{(Color online) Distributions of (a) off times, (b) on times, and (c) lengths $L=\int_{t_{on}}^{t_{off}}v(t)dt$. Blue curves correspond to results based on $v(t)$ from particle imaging experiments, in which 1436 events were observed. Red curves correspond to $\delta v(t)$ from speckle-visibility experiments, in which 1472 events were observed. The averages are $\langle T_{\rm off} \rangle_{\delta v} = 60.0$~s, $\langle T_{\rm off} \rangle_{v} = 62.3$~s, $\langle T_{\rm on} \rangle_{\delta v} = 14.9$~s, $\langle T_{\rm on} \rangle_{v} = 16.6$~s, and $\langle L \rangle =44.8$ cm.}
\label{statistics}
\end{figure*}

\begin{figure*}
\includegraphics[width=6.5in]{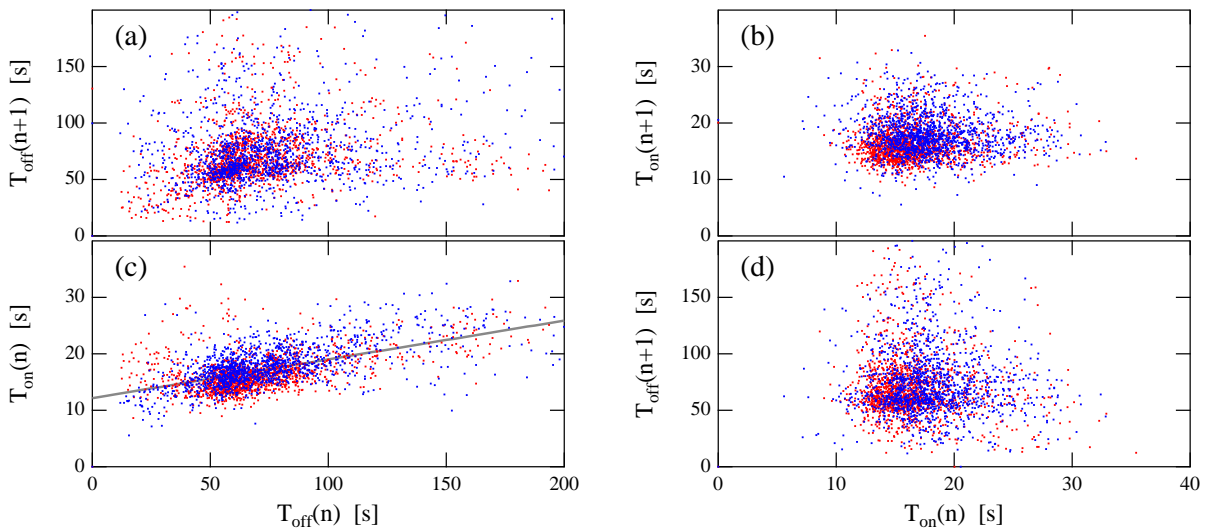}
\caption{(Color online) Scatter plots of on and off times for successive events, as labeled.  As defined in Fig.~\protect{\ref{vtdvt}}, the temporal sequence is $T_{\rm off}(n)$, $T_{\rm on}(n)$, $T_{\rm off}(n+1)$, $T_{\rm on}(n+1)$, etc. Significant correlation exists only in part-c, as indicated by the solid gray line.} \label{scatter}
\end{figure*}

Histograms for the rest time $T_{\rm off}$ between avalanches, and for the duration $T_{\rm on}$ of events, are collected in Figs.~\ref{statistics}a-b.  Whether measured by imaging or SVS, the results show that the distribution of rest times is strongly peaked with an average of $\langle T_{\rm off}\rangle=61$~s and a standard deviation of 34~s.  The minimum rest time is about 10~s, which is 1.5~standard deviations below the average; therefore successive avalanches are well-separated and quasi-static.  As for the rest times, the distribution of event durations is also strongly peaked, with an average of $\langle T_{\rm on}\rangle=16$~s and a standard deviation of 4~s.  While both distributions are skewed toward longer times, they have a rapid final decay inconsistent with a power-law.  Hence the events are quasi-periodic; the average period is $\langle T_{\rm off}+T_{\rm on}\rangle=77~{\rm s}$.

A length scale quantifying event size may be defined by integrating the flow speed over the event duration, $L=\int v(t){\rm d}t$.  The distribution of event lengths, shown in Fig.~\ref{statistics}c, is once again peaked with an average of $\langle L\rangle = 45$~cm and a standard deviation of 14~cm.  The histograms for $L$ and $T_{\rm on}$ have slightly different shapes because $v(t)$ decreases throughout the event.  Note that $\langle L\rangle$ is about one standard deviation longer than the distance along the slope of the heap, consistent with avalanches that sweep through the entire system.  Further note that the average thickness of the flowing layer is given by mass conservation as $Q \langle T_{\rm on}+T_{\rm off}\rangle / (\rho \langle L\rangle w)$, which equals about three grain diameters - consistent with visual observation and thinning of the layer during the course of an event.

Next we consider possible correlations between the rest times and event sizes. As in the example time trace, Fig.~\ref{vtdvt}, the event index $n$ is defined such that $T_{\rm off}(n)$ is the rest time immediately prior to the $n^{\rm th}$ avalanche, which has duration $T_{\rm on}(n)$.  Scatter plots of flow and rest times vs preceding rest time, and of rest and flow times vs preceding flow time, are displayed in Fig.~\ref{scatter}a-d.  The only noticeable correlation is in part-c, which shows that $T_{\rm on}(n)$ grows with increasing $T_{\rm off}(n)$.  Intuitively, for longer rest times more grains accumulate at the top of the heap and so the next event is larger.  There appears to be no such correlation between successive events.  Thus $T_{\rm off}$ is determined at random from the distribution shown in Fig.~\ref{statistics}, but $T_{\rm on}$ is not.

\subsection{Switching probabilities}

\begin{figure}
\includegraphics[width=3.0in]{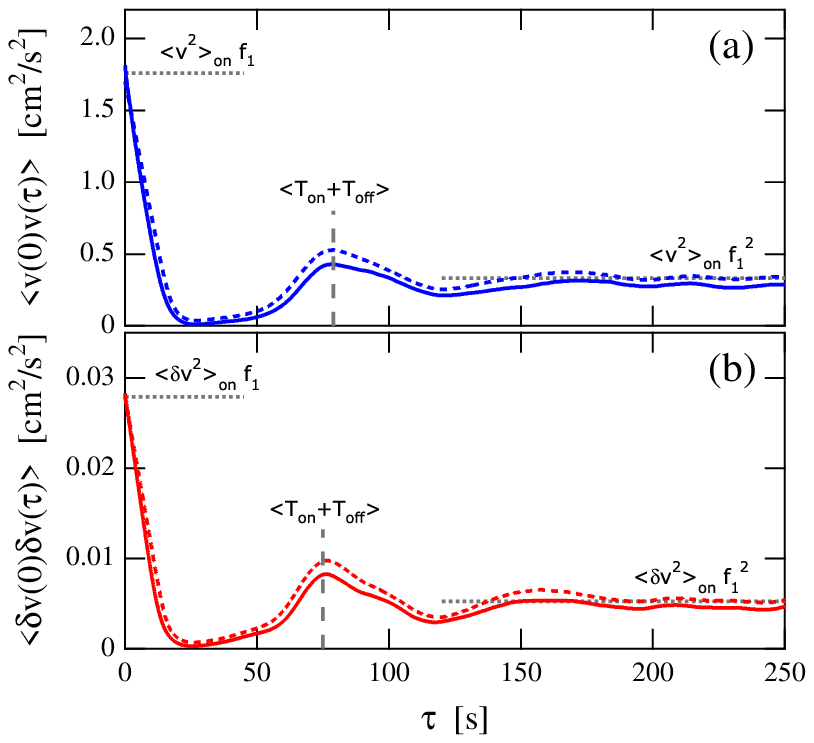}
\caption{(Color online) Autocorrelation functions of (a) $v(t)$ and $\delta v(t)$ data.  The average values $\langle T_{\rm on} \rangle$, $\langle T_{\rm off} \rangle$, and $f_1 = \Sigma T_{\rm on}/(\Sigma T_{\rm on}+\Sigma T_{\rm off}$) are computed from distributions in Fig.~\ref{statistics}. Solid curves are autocorrelations of actual data, while dashed curves are autocorrelations of the telegraph-approximated signals.} \label{autocorrelations}
\end{figure}

\begin{figure}
\includegraphics[width=3.0in]{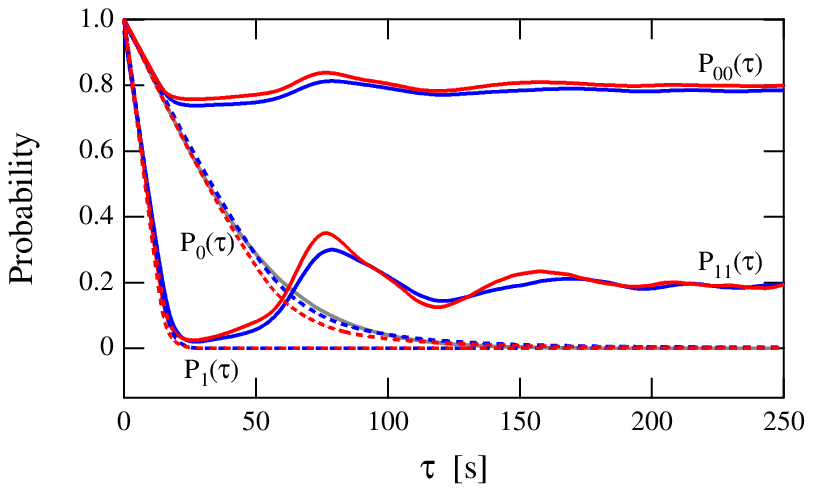}
\caption{(Color online) Probability functions for switching between states vs delay time, as labeled.  Blue and red curves correspond to results based on $v(t)$ and $\delta v(t)$ data, respectively.  The solid gray curve is $[1 + 3(\tau/T_{\rm off}) + 4(\tau/T_{\rm off})^2 + (8/3)(\tau/T_{\rm off})^3]\exp(-4\tau/T_{\rm off})$.} \label{probs}
\end{figure}

To study the dynamics by which avalanches start and stop, we begin by computing the autocorrelation of the $v(t)$ and $\delta v(t)$ time trace data.  The results, plotted in Figs.~\ref{autocorrelations}a-b, display a linear decay followed by damped oscillations to a constant.  This is consistent with quasi-periodic behavior.  Note that the initial decay time is set by $\langle T_{\rm on}\rangle$ and that the first peak occurs at the period $\langle T_{\rm off}+T_{\rm on}\rangle$.  The detailed shape of the speed autocorrelations is determined both by the variation of the speed during an event, as well as by the statistics by which the flow starts and stops.  To investigate the relative importance of these contributions, we compare with expectation based on telegraph-approximated signals $v(t) \approx \sqrt{\langle v^2\rangle} x(t)$ and $\delta v(t) \approx \sqrt{\langle \delta v^2 \rangle} x(t)$, where $x(t)$ is equal to 0 during rest and is equal to 1 during flow.  The resulting autocorrelations, displayed in Figs.~\ref{autocorrelations}a-b, have nearly the same shape as the actual speed autocorrelations.  Note that the initial and final expectations are set by the mean-squared speeds multiplied by $\langle x^2\rangle=f_1$ and $\langle x\rangle^2={f_1}^2$, respectively.  The good agreement between actual and telegraph-approximated autocorrelations implies that the variation of speed during an event has only minor consequence.  This supports the validity of the analysis of the higher-order intensity correlation data presented in Refs.~\cite{PierrePRL00, PierreAO01}.

To fully characterize the dynamics of switching between flow and rest states, we employ standard probability functions as in Refs.~\cite{PierrePRL00, PierreAO01}.  Thus we define $P_{ij}(\tau)$ as the conditional probability for the system to be in state $j$ at time $t+\tau$ if it started in state $i$ at time $t$.  By convention subscript 0 denotes a state of rest, and subscript 1 denotes at state of flow.  All four of these inter-related functions may be computed from the speed vs time data, in terms of the autocorrelation $\langle x(0)x(\tau)\rangle$ of the corresponding telegraph signals.  The relevant identities are $\langle x(0)x(\tau) \rangle = f_1P_{11}(\tau)$, $P_{00}(\tau) + P_{01}(\tau) = 1$, $P_{10}(\tau) + P_{11}(\tau) = 1$, and $f_0 P_{01}(\tau) = f_1 P_{10}(\tau)$.  In addition we define two more functions, $P_0(\tau)$ and $P_1(\tau)$, as the probabilities to be in the {\it same} on or off state at time $t+\tau$ as at time $t$, with no changes of state in between.  These may be computed by averaging over all off times as $P_0(\tau) = \langle [1-\tau/T_{\rm off}(n)] H(T_{\rm off}(n)-\tau) \rangle$, and similarly for $P_1(\tau)$, where $H(t)$ is the Heaviside function.  The function $P_0(\tau)$ is particularly crucial for analysis of dynamic light scattering data \cite{PierrePRL00, PierreAO01}.

Results for the switching probabilities are collected in Fig.~\ref{probs}.  The initial decays of $P_0(\tau)$ and $P_{00}(\tau)$ are both linear, $1-\tau/T_{\rm off}$.  While the former decays fully to zero, since all rest states have finite duration, the latter oscillates and decays to $f_0=0.8$ due to contributions from other rest states.  Note that the avalanches are quasiperiodic, in that about three oscillations occur before the full decay to $f_0$.  Similar statements hold for the analogous flowing state switching functions.  The functional form for $P_0(\tau)$ is similar to that found in Ref.~\cite{PierreAO01} using multiple light scattering.  Namely, it is faster than exponential and well-described by $[1 + 3(\tau/T_{\rm off}) + 4(\tau/T_{\rm off})^2 + (8/3)(\tau/T_{\rm off})^3]\exp(-4\tau/T_{\rm off})$.  This function is plotted as a solid gray curve; its limiting behavior is $1-(\tau/T_{\rm off})+(32/15)(\tau/T_{\rm off})^5+O(\tau/T_{\rm off})^6$.

\subsection{Average event dynamics}

\begin{figure*}
\includegraphics[width=6.5in]{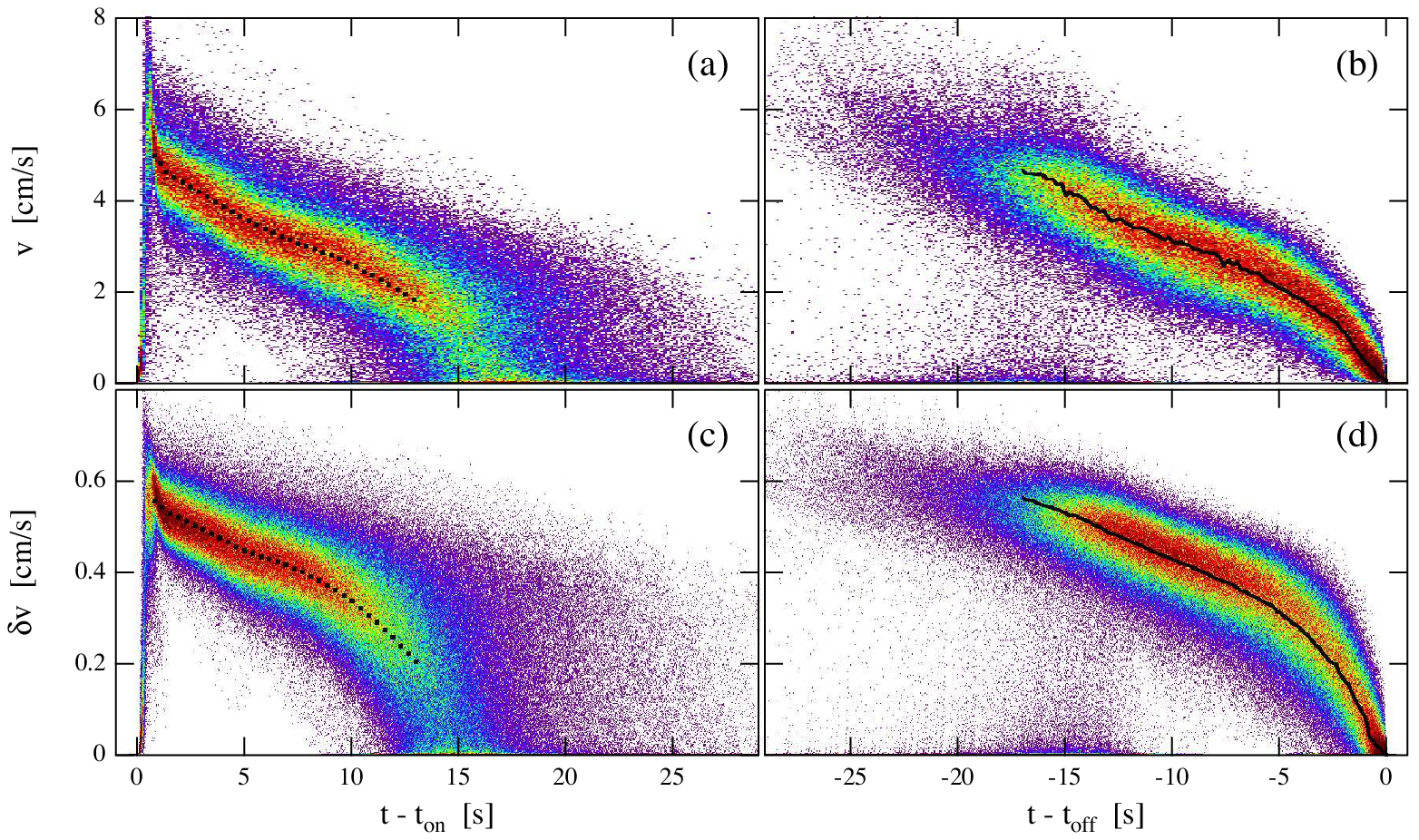}
\caption{(Color online) Probability density maps (rainbow colors), and most probable avalanche dynamics (black curves), for the average speed $v(t)$ and fluctuation speed $\delta v(t)$ vs time $t$ during an avalanche.  All observed avalanches are included by lining up their individual time traces according to either when they turn on (a,c) or else according to when they turn off (b,d).  The bin sizes are $0.02$~cm/s $\times$ $0.1$~s for $v(t)$ data and $0.002$~cm/s $\times$ $1.72$~ms for $\delta v(t)$ data.} \label{shapes}
\end{figure*}

\begin{figure}
\includegraphics[width=3.0in]{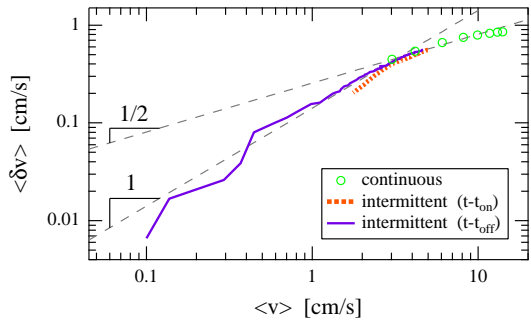}
\caption{(Color online) The relation between $\langle \delta v \rangle$ and $\langle v\rangle$ for both continuous flows (open green circles) and avalanching flows (curves).  The dashed orange curve represents $\langle \delta v(t-t_{on}) \rangle$ vs $\langle v(t-t_{on}) \rangle$, based on the dashed curves in Fig.~\protect{\ref{shapes}}a and c.  The solid purple curve represents $\langle \delta v(t-t_{off}) \rangle$ vs $\langle v(t-t_{off}) \rangle$, based on the solid curves in Fig.~\protect{\ref{shapes}}b and d.}
\label{vdvrelation}
\end{figure}

In this final section we consider the dynamics of individual avalanches, and how the average and fluctuation speeds evolve during the course of an event.  Since there is a range of avalanche sizes and durations, the results are shown in Fig.~\ref{shapes} as probability densities for there to a given speed at a given time.  In parts (a,c), the typical behavior at the beginning of an event is revealed by binning the speed data for all events vs the time $t-t_{\rm on}$ since the start of flow.  The initial rise from zero is quick, set by the sweep of a moving front through the field of view.  Once the speeds reach a maximum, the average results for $v(t)$ and $\delta v(t)$ decrease nearly linearly with time during a span of about ten seconds.  This is indicated by the dashed black curves, which trace along the crest of the probability distributions.  After 10-15~s, the typical behavior becomes less well-defined since some events stop and some continue.  In parts (b,d), the typical behavior at the end of an event is revealed by binning the speed data for all events vs the time $t-t_{\rm off}$ until the avalanche ceases.  As before, the crest of the distributions is indicated by a black curve.  The data show that the speeds continue the initial linear decrease with time up until about 5~s before cessation.  During the final stage, both $v(t)$ and $\delta v(t)$ decrease ever more rapidly with time until reaching zero.  While there is no discontinuity in speeds at $t=t_{\rm off}$, the slopes vanish abruptly.

To examine the relationship between flow and fluctuation speeds, we display a log-log parametric plot of $\langle \delta v(t)\rangle$ vs $\langle v(t)\rangle$ in Fig.~\ref{vdvrelation}.  Since PIV and SVS data were collected separately with the same camera, we do not have simultaneous flow and fluctuation speed data for the same avalanches.  Therefore, the results on display are for a typical event given by the black curves in Fig.~\ref{shapes} tracing along the crests of the probability distributions.  In other words, $\langle \dots \rangle$ represents the average taken over an ensemble of events.  When events are aligned according to start time, $t-t_{on}$, the relation between $\langle \delta v(t)\rangle$ vs $\langle v(t)\rangle$ is displayed as a orange dashed curve.  And when events are aligned according to stop time, $t-t_{off}$, the relation between $\langle \delta v(t)\rangle$ vs $\langle v(t)\rangle$ is displayed as a solid purple curve.  These two curves agree well, but the latter has a larger range since both speeds vanish at the end of the event.  The relation is approximately $\langle \delta v \rangle = 0.1 \langle v\rangle$, as shown by the dashed line with slope 1; during an an avalanche event, the fluctuation speed is proportional to the flow speed, but is ten times smaller.  For smooth continuous flow at high flux, $Q>Q_c$, a different relation is found.  The data, shown by open green circles in Fig.~\ref{vdvrelation}, are consistent with $\langle \delta v \rangle \sim \sqrt{\langle v\rangle}$ as shown by the dashed line with slope $1/2$.  Similar behavior was observed in hopper flow, where the exponent was approximately $2/3$ \cite{Menon1997}.  For continuous flows, the fluctuation speed decreased more gradually than the average flow speed as the flux is lowered.  This results in proportionately greater dissipation, and hence a transition to intermittent flow at nonzero $Q_c$.

It is curious to note in Fig.~\ref{vdvrelation} that $\langle \delta v\rangle$ vs $\langle v\rangle$ data for continuous and intermittent regimes are not disjoint.  At the beginning of an avalanche, and for continuous flow at fluxes slightly greater than $Q_c$, the average flow speeds can be the same.  Across this overlap, the fluctuation speeds in the two regimes are in good agreement.  Therefore, the nature of granular heap flow at the beginning of an avalanche is remarkably similar to that for continuous flow at low flux.  In this sense, there is a smooth crossover from $\langle \delta v\rangle \sim \langle v\rangle$ for avalanches to $\langle \delta v\rangle \sim \sqrt{\langle v\rangle}$ for continuous flow.

%=========================================================================================
\section{Conclusion}

Speckle-Visibility Spectroscopy has permitted us to observe the instantaneous velocity fluctuations of sand particle for continuous and avalanching flows over a large dynamical range.  Particle-image velocimetry has allowed us to observe the instantaneous flow velocity of sand particles in continuous and avalanching flows over a similarly large dynamical range. Together, these observation methods provide $\delta v$ and $v$ measurements that allow us to significantly improve upon previous studies~\cite{PierrePRL00, PierreAO01} and to thoroughly characterize both the macroscopic and microscopic dynamics of avalanches.

The ability to observe thousands of distinct avalanches over the course of 74 hours of observation has permitted us to uncover the full time-on and time-off distributions. We learn that even for constant sand addition flow rates, avalanches come in a wide variety of sizes, and the on and off times of avalanches are sampled from distributions that are peaked but non-Gaussian. The avalanches are quasi-periodic over only a few avalanche cycles, and the flow periodicity is essentially decorrelated after three cycles.

The ability to fully observe the instantaneous flow and fluctuation dynamics of independent avalanches has allowed us to uncover their rich and characteristic dynamical shape. There is a sharp wavefront at the head of the avalanche, continuous deceleration in the middle, and abrupt cessation of flow at the end. Thus, even though the simple square wave of the telegraph model works well on long timescales, it is a poor approximation to the richly detailed shape of individual avalanches in the flowing state.

Finally, because we measure $\delta v$ and $v$ instantaneously for tens of hours and thousands of avalanches, we are able to directly test the validity and range of applicability of previously reported functional forms. The result is that we learn that the functional form $\delta v \sim v^{1/2}$ is only correct for highly fluidized continuous granular flows. For intermittent
flows, this law progressively fails and gives way to $\delta v \sim v$ precisely at the critical flow rate crossover $Q_c$ that separates continuous from avalanching flows.

These results combine to give a detailed picture of avalanches at both the microscopic and macroscopic levels and with the range necessary to fully characterize dynamics. The microscopic dynamics of sand particles depend on the macroscopic flow state of the pile. For example, the ``granular temperature'' in a continuously decelerating avalanche is lower than would be expected by extrapolating back from the continuous flow relation $\delta v \sim v^{1/2}$, in which correlated particle collisions are a necessary ingredient~\cite{Menon1997}. Instead, microscopic dynamics of intermittent avalanches fit far better to the form $\delta v \sim v$. New theoretical insights may uncover how changing flow-structure of decelerating avalanches~\cite{duPont2005} affects the collision rules of sand particles at the microscopic level, which are empirically found to be different on either side of the critical transition point $Q_c$.

\begin{acknowledgments}
We thank S.S. Suh for helpful discussions. Our work was supported by the National Science Foundation through grant DMR-0704147, by the Aspen Center for Physics, and by the Japanese Society for the Promotion of Science through a Postdoctoral Fellowship for Research Abroad (HK).
\end{acknowledgments}

% figure list
% 1 vraw        (large)
% 2 dvraw       (large)
% 3 vtdvt
% 4 statistics
% 5 scatter
% 6 autocorrelations
% 7 probs
% 8 shapes      (large)
% 9 vdvrelation
%
% To reduce eps size:
% [1] open with Ghostscript and convert/save as PDF 300
% [2] open with Adobe and save as PS
% [3] open with Ghostscript and Convert PS to EPS
s

\bibliography{avalanche_refs}

\end{document}